\documentclass[12pt]{spieman}  
\usepackage{amsmath,amsfonts,amssymb}
\usepackage{graphicx}
\usepackage{setspace}
\usepackage{tocloft}
\usepackage{verbatim}
\usepackage{hyperref}
\usepackage{xcolor}

\usepackage{lineno}

\title{Modeling human observer detection in undersampled magnetic resonance imaging (MRI) reconstruction with total variation and wavelet sparsity regularization}

\author[a]{Alexandra G. O'Neill}
\author[a]{Emely L. Valdez}
\author[b]{Sajan Goud Lingala}
\author[a,*]{Angel R. Pineda}
\affil[a]{Manhattan College, Department of Mathematics, 4513 Manhattan College Pkwy, The Bronx, NY, USA, 10471}
\affil[b]{University of Iowa, Roy J. Carver Department of Biomedical Engineering, Iowa City, IA, USA, 52242}

\cftpagenumbersoff{figure}
\cftpagenumbersoff{table} 
\begin{document} 
\maketitle

\begin{abstract}

\noindent {\bf Purpose:} Task-based assessment of image quality in undersampled magnetic resonance imaging (MRI) provides a way of evaluating the impact of regularization on task performance. In this work, we evaluated the effect of total variation (TV) and wavelet regularization on human detection of signals with a varying background and validated a model observer in predicting human performance.

\noindent {\bf Approach:} Human observer studies used two-alternative forced choice (2-AFC) trials with a small signal known exactly (SKE) task but with varying backgrounds for fluid-attenuated inversion recovery (FLAIR) images reconstructed from undersampled multi-coil data.  We used a 3.48 undersampling factor with TV and a wavelet sparsity constraints.  The sparse difference-of-Gaussians (S-DOG) observer with internal noise was used to model human observer detection. The internal noise
for the S-DOG was chosen to match the average percent correct (PC) in 2-AFC studies for four observers
using no regularization.  That S-DOG model was used to predict the percent correct of human observers for a range of regularization parameters.

\noindent {\bf Results:} We observed a trend that the human observer detection performance remained fairly constant for a broad range of values in the regularization parameter before decreasing at large values. A similar result was found for the normalized ensemble root mean squared error (ERMSE).   Without changing the internal noise, the model observer tracked the performance of the human observers as the regularization was increased but over-estimated the PC for large amounts of regularization for TV and wavelet sparsity, as well as the combination of both parameters. 

\noindent {\bf Conclusions:} For the task we studied, the S-DOG observer was able to reasonably predict human performance with both total variation and wavelet sparsity regularizers over a broad range of regularization parameters.  We observed a trend that task performance remained fairly constant for a range of regularization parameters before decreasing for large amounts of regularization.
\end{abstract}


{\noindent \footnotesize\textbf{*}Angel R. Pineda,  \linkable{angel.pineda@manhattan.edu} }

\begin{spacing}{2}   

\section{Introduction}
\label{sect:intro}
Task-based assessment of image quality \cite{Barrett1990,BM04} for reconstructed MRI images is critical to the development and validation of accelerated reconstruction techniques.  This approach has been extensively used in other imaging modalities \cite{McCollough2017, Favazza2017, Kupinski2016, Grace2011,Gifford2007, Brankov_2013, baek2013bin, jha2013, Pineda2006ODT,Pineda2006}  There has been research using task-based techniques in MRI for parallel imaging \cite{Wilson2006} and compressed sensing \cite{Graff2015} but most methods utilized for assessment of image quality in MRI are typically root mean square error (RMSE) and structural
similarity index (SSIM). These are measures of pixel value differences between the original and reconstructed image\cite{Wang2004,Lustig2007}. While these measures do give an indication of similarity between images, neither RMSE nor SSIM take into account the specific task for which the image will be used. As a result, images with the same RMSE and SSIM could produce different performance in these tasks. Observer models are an alternative way to assess image quality by taking into account human visual principles as well as the task for which the image will be used \cite{abbey2001human, burgess,myers1987addition}.

Previous results related to evaluating MRI reconstruction of undersampled images using an approximation to the ideal linear observer \cite{Chen2017,pinedaLG, PinedaPMB2021} showed that the area under the ROC curve (AUC) for the channelized Hotelling observer with Laguerre Gauss channels showed only a small improvement with regularization. However, both SSIM and RMSE showed a large improvement. Our preliminary results using human observer studies \cite{ONeill2021, Oneill2022}  suggest a similar result in that there was no large improvement with regularization for this task. The purpose of this work is to evaluate constrained reconstruction of undersampled MRI data using TV, wavelet, and a combination of these constraints based on human observer performance in detecting a small signal in a 2-AFC task. A model observer (S-DOG) is used to model human observer performance in this detection task. Along with optimizing reconstruction, a goal of this work is also to reduce the number of future human observer studies needed in undersampled MRI reconstruction by using model observers to evaluate image quality.

\section{Methods}

\subsection{Undersampled acquisition in MRI}
\begin{figure} [ht]
   \begin{center}
   \begin{tabular}{c} 
   \includegraphics[width=.9\linewidth]{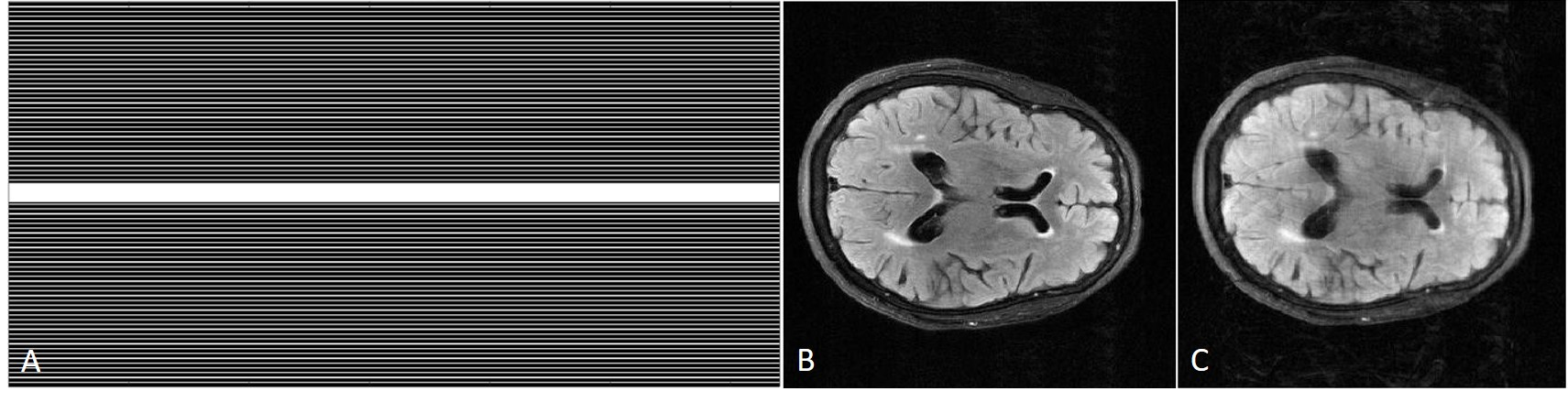}
   \end{tabular}
   \end{center}
   \caption[(A) Sampling mask for 4x undersampling, (B) Fully sampled image, (C) Undersampled constrained reconstruction with no TV regularization,  Aliasing in the vertical axis is visible in the undersampled image. The acquisition had a 2x oversampling in the horizonal direction. ]
   { \label{fig:Acquistion}(A) Sampling mask for 4x undersampling, (B) Fully sampled image, (C) Undersampled constrained reconstruction with no TV regularization,  Aliasing in the vertical axis is visible in the undersampled image. The acquisition had a 2x oversampling in the horizonal direction.}
   \end{figure}

For this study we consider 1-D undersampling of FLAIR images (Figure \ref{fig:Acquistion}) with an acquisition that samples every fourth phase encoding line plus fully sampling the middle 16 kspace lines resulting in an effective undersampling factor of 3.48. Data used in the preparation of this article were obtained from the NYU fastMRI Initiative database \cite{knoll}.  The average white matter signal to noise ratio (SNR) for the fully sampled multi-coil SENSE reconstructions (R=1) was measured by estimating the standard deviation from components of the image with only noise and assuming a Rayleigh distribution \cite{Dietrich2007} and the mean signal across a homogeneous region of white matter image in the reconstructed slices.  This led to an average white matter SNR of 83 in the 50 slices used for the observer studies.  The NYU fastMRI investigators provided data but did not participate in analysis or writing of this report. 

\subsection{Constrained Reconstruction from multi-coil Data}
\label{sect:title}
\begin{figure} [ht]
   \begin{center}
   \begin{tabular}{c} 
   \includegraphics[width=.9\linewidth]{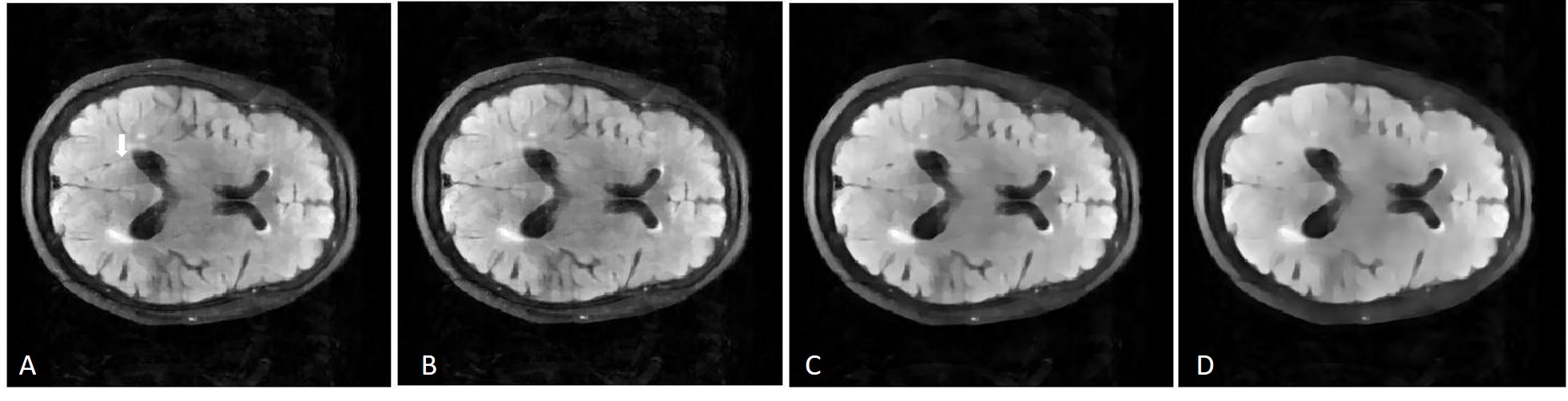}
   \end{tabular}
   \end{center}
   \caption[Sample undersampled images reconstructed with TV regularization, A) $\alpha_{TV}=0.01$, B) $\alpha_{TV}=0.02$, C) $\alpha_{TV}=0.05$, D) $\alpha_{TV}=0.1$.  As the regularization increases, there are reduced aliasing artifacts but also reduced resolution.  The arrow in image A shows the location of one of the undersampling artifacts.] 
   { \label{fig:TVImages} Sample undersampled images reconstructed with TV regularization, A) $\alpha_{TV}=0.01$, B) $\alpha_{TV}=0.02$, C) $\alpha_{TV}=0.05$, D) $\alpha_{TV}=0.1$.  As the regularization increases, there are reduced aliasing artifacts but also reduced resolution.  The arrow in image A shows the location of one of the undersampling artifacts.
}
   \end{figure}

\begin{figure} [ht]
   \begin{center}
   \begin{tabular}{c} 
   \includegraphics[width=.9\linewidth]{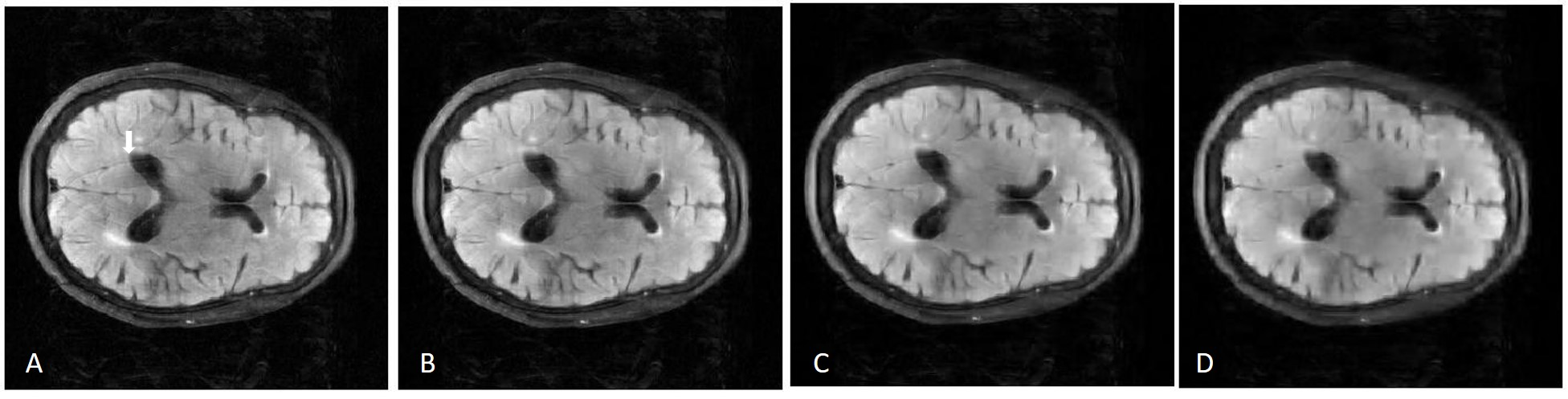}
   \end{tabular}
   \end{center}
   \caption[Sample undersampled images reconstructed with wavelet sparsity, A) $\alpha_{W}=0.01$, B) $\alpha_{W}=0.02$, C) $\alpha_{W}=0.05$, D) $\alpha_{W}=0.1$.  As the regularization increases, there are reduced aliasing artifacts but also reduced resolution but with different texture than in TV regularization.  The arrow in image A shows the location of one of the undersampling artifacts.] 
   { \label{fig:WaveletImages} Sample undersampled images reconstructed with wavelet sparsity, A) $\alpha_{W}=0.01$, B) $\alpha_{W}=0.02$, C) $\alpha_{W}=0.05$, D) $\alpha_{W}=0.1$.  As the regularization increases, there are reduced aliasing artifacts but also reduced resolution but with different texture than in TV regularization.  The arrow in image A shows the location of one of the undersampling artifacts.
}
   \end{figure}

Constrained reconstruction minimizes a data agreement functional with additional constraints. For this
work we consider a total variation and wavelet constraint\cite{Lustig2007} and multi-coil parallel imaging using SENSE\cite{PruessmannSENSE1999} with the coil sensitivities
estimated using the sum of squares method which leads to real estimates of real-valued objects.\\
The following functional was minimized when reconstructing these images:
\begin{equation}
||H(f)-g||^2_2 + \alpha_{\Psi} \, ||\Psi(f)||_1 + \alpha_{TV} \, ||\nabla f||_1,
\end{equation}
\noindent where $H(f)$ is the undersampled Fourier operator acting on the discretized object $f$, the Fourier data is ${\bf g}$, $\Psi$ is the Daubechies 4 wavelet transform, $||\nabla f||_1$ is the $L_1$ norm of the gradient which is the total variation operator and $\alpha_{\Psi}$ and $\alpha_{TV}$ are the regularization parameters for the wavelet and TV constraints respectively \cite{Lustig2007}.  Sample images with different amounts of TV and wavelet regularization are shown in Figure \ref{fig:TVImages} and Figure \ref{fig:WaveletImages} respectively.  The reconstruction of the images was done using
the Berkeley Advanced Reconstruction Toolbox (BART) toolbox \cite{uecker}.   The ensemble root mean squared error for the reconstructed images (ERMSE) was computed using the fully sampled unregularized reconstruction normalized to [0,1] as the reference image.  The underampled reconstructions were normalized to [0,1] before computing the RMSE.  The ERMSE was computed from 50 slices from 5 volumes.


\subsection{Two-alternative forced choice experiments}
\begin{figure} [ht]
   \begin{center}
   \begin{tabular}{c} 
   \includegraphics[width=.7\linewidth]{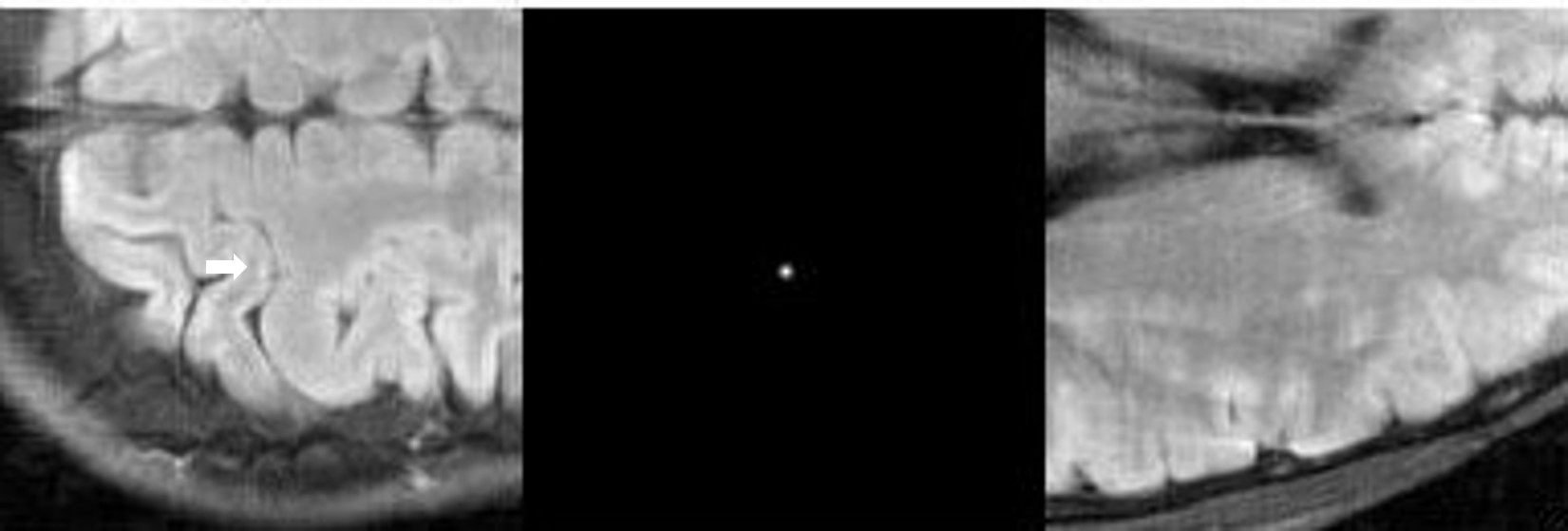}
   \end{tabular}
   \end{center}
   \caption[Sample 2AFC trial with signal in the left image.  The arrow present in this image was not used in the observer study and is there to help the reader identify the signal.  In order to see differences in performances due to regularization, the images used were subtle.  These images were generated with no regularization but the artifacts are more visible in the image to the right. ] 
   { \label{fig:2AFC} Sample 2AFC trial with signal in the left image.  The arrow present in this image was not used in the observer study and is there to help the reader identify the signal.  In order to see differences in performances due to regularization, the images used were subtle.  These images were generated with no regularization but the artifacts are more visible in the image to the right. 
}
   \end{figure}

In each individual trial of the 2AFC experiment we presented three 128x128 pixel images: one image of an anatomical background with the signal, the signal, and one image of an anatomical background without the
signal. The signal image was always in the center, and the location (left or right) of the anatomical image
with the signal was randomly chosen for each trial.   All images are scaled to [0,1] before being displayed using the 8 bit gray scale colormap in MATLAB.  Since the observers were not radiologists, the images were windowed and leveled for them.  Our choice of windowing leads to a different number of gray scale values representing the signals depending on the backgrounds and regularization.  In all cases, a reasonable number of gray scale values was used to represent the signal contrast with an average of 32 gray scale values for images with no regularization down to an average of 8 gray scale values for images with the most regularization ($\alpha_{TV}$=0.1, $\alpha_{W}$=0.1).  An example trial is shown in Figure \ref{fig:2AFC}. The signal location is always
in the middle of the anatomical image, which makes this task a signal known exactly (SKE) with varying backgrounds and the human observer only determines whether or not it is present in the image. 
   
\subsection{Experimental procedure}
For each experimental condition, 200 2AFC trials were carried out by four observers. As training, all
observers repeated an initial set of 200 trials until the performance plateaued. Based on performance in the training trials, the signal amplitude was
chosen so that the mean percent correct for the observers would be close to 90 percent when no regularization was used. 

The observer studies were done using a Barco MDRC 2321 monitor in a dark room. The resolution of the Barco monitor was 0.294 mm and the observers were approximately 50 cm from the screen.  A marker at 50 cm was provided for reference.

For each individual trial, a set of images like Figure \ref{fig:2AFC} appeared on the screen and the observer chose which image they believed contained the signal. The observers received feedback on
whether they had identified the signal correctly after each trial for possible improvement between trials and took breaks between sets of 200 trials to avoid fatigue.

For each observer, the images used in each 2-AFC trial were randomly  paired from 200 backgrounds with the signal and 200 background images.  For each level of regularization, the standard deviation was computed using the percent correct of each of the four observers.  There are several sources of variability in this estimate of the percent correct due to the multiple reader - multiple case (MRMC) experimental design.  Our estimate of the standard deviation based on the percent correct of the individual observers is  a summary of the variability of the percent correct values from the readers but not based on an unbiased estimate \cite{Gallas2007}.

\subsection{Model Observers}

\begin{figure} [ht]
   \begin{center}

 \includegraphics[width=1\linewidth]{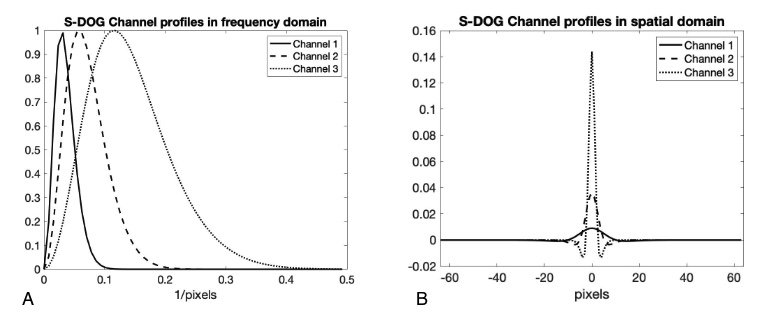}

   \end{center}
   \caption[(A) The cross section of S-DOG channels matched for human performance with no TV regularization in the frequency domain and (B) spatial domain.] 
   { \label{fig:S-DOGchannels} (A) The cross section of S-DOG channels matched for human performance with no TV regularization in the frequency domain and (B) spatial domain.}
   \end{figure}

The model that we used in our study was the S-DOG channelized Hotelling Observer \cite{abbey2001human}.  The channel data ${\bf g_c}$ is obtained by taking the inner product of the image with the channels centered at the signal. The elements $g_{c,j}$ of the vector ${\bf g_c}$ are given by:
\begin{equation}
g_{c,j} = C_j^t f,
\end{equation}
where $C_j$ is the $j^{th}$ channel in the spatial domain and $f$ is the object in the spatial domain.
We used the S-DOG channels of the form 
\begin{equation}
     C_{j}(k) = exp[-\frac{1}{2}\left(\frac{k}{Q*\sigma_{j}}\right)^2]- exp[-\frac{1}{2}\left(\frac{k}{\sigma_{j}}\right)^2]
\end{equation}
in the frequency domain where $k$ is the distance from the zero frequency, Q is the multiplicative factor of the bandwidth,  $\sigma_{j} = \sigma_{0}*\alpha^{j} $, where j denotes the j$^{th}$ channel and $\sigma_{j}$ denotes the standard deviation of each channel. The parameters that were used for the S-DOG were $Q = 2$,
$\alpha= 2$, and $\sigma_0= 0.015$ which were used by Abbey\cite{abbey2001human} (Figure \ref{fig:S-DOGchannels}).

The test statistic ($\lambda$) for the S-DOG observer for each image was computed using:
\begin{equation}
\lambda = [(K_c + K_{int})^{-1}\bf{s_c}]^t (\bf{g_c} + \bf{n_{int}}),
\end{equation}
where $K_c$ is the sample covariance matrix of the data in the channel domain, $K_{int}$ is the covariance matrix of the channel internal noise, $s_c$ is the mean signal difference in the channel domain, $g_c$ is the image in the channel domain and $n_{int}$ is a noise vector drawn from the distribution $N(0,K_{int})$.

The covariance for the internal noise is:
\begin{equation}
K_{int} = \epsilon \, diag(K_c),
\end{equation}
where $\epsilon$ is the internal noise constant and $diag(K_c)$ is a diagonal matrix with the diagonal elements of $K_c$ \cite{abbey2001human}.

\noindent The channel covariance matrix $K_c$ is computed by taking the average of the channel covariance matrix with and without the signal.  The mean signal difference ($s_c$)is computed by subtracting the sample average of the channel outputs without the signal from the average channel outputs with the signal.  The internal noise was determined by varying the noise constant $\epsilon$ values until the performance matched the average human performance for the unregularized reconstruction.
 This calibrated model with the images with no regularization was used to predict all other experimental conditions with regularization.  
   
We utilized six regularization parameter values for each of the three regularization conditions that were generated with 4x undersampling. For TV and wavelet regularization, the parameters used were: 0, 0.001, 0.002, $2*10^{-1}$, $5*10^{-1}$, and 0.1. These values were also used when both types of regularization were combined.

\section{Results and Discussion}

\begin{figure} [ht]
   \begin{center}
   \includegraphics[width=\linewidth]{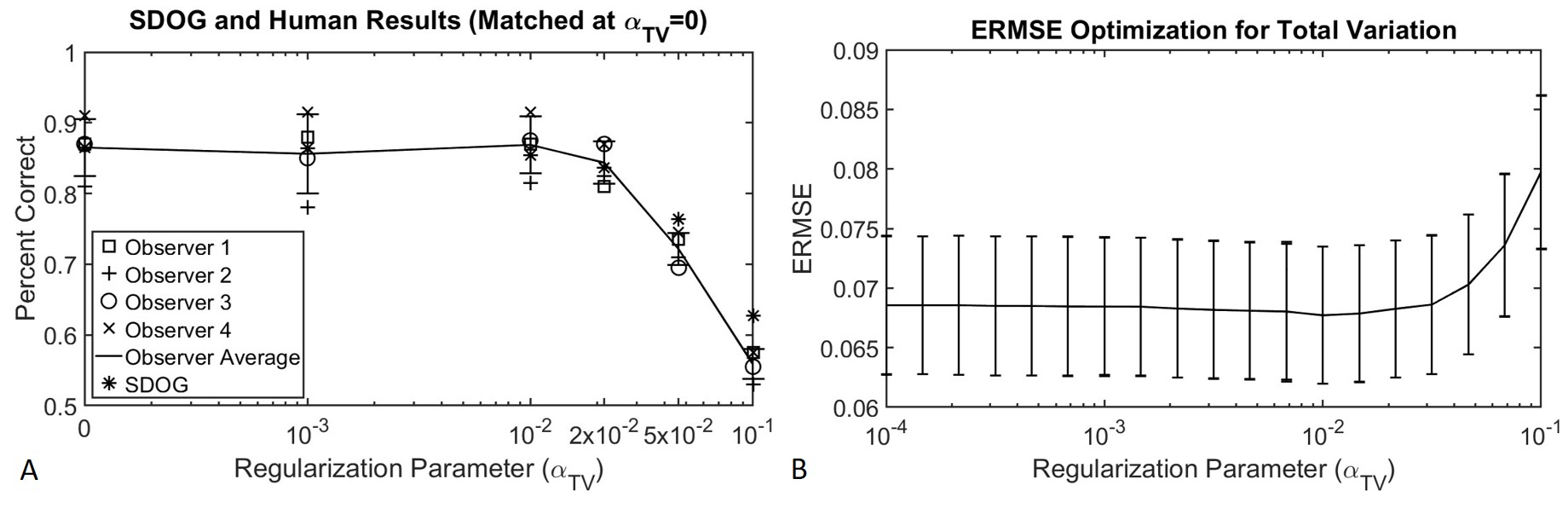}

   \end{center}
   \caption[A) S-DOG observer with internal noise matched to average human observer for images with no TV regularization. The S-DOG observer tracked the human observer data for the other regularization parameters. The S-DOG slightly overestimates for the largest regularization parameters.  B) Normalized ERMSE for varying TV regularization.  The ERMSE metric leads to similar conclusions as the 2-AFC detection performance.] 
   { \label{fig:S-DOGfitTV} A) S-DOG observer with internal noise matched to average human observer for images with no TV regularization. The S-DOG observer tracked the human observer data for the other regularization parameters. The S-DOG slightly overestimates for the largest regularization parameters.  B) Normalized ERMSE for varying TV regularization.  The ERMSE metric leads to similar conclusions as the 2-AFC detection performance.}
\end{figure}

The S-DOG overestimated human performance without the addition of internal noise. However, the pattern of performance as TV increased was the same as both the human observers.  Once the internal noise was added, tracking the human performance can be seen with the S-DOG in Figure \ref{fig:S-DOGfitTV}. The S-DOG closely tracks human performance for small amounts of regularization (which would be used in practice) and over-estimates the human performance for large amounts of regularization.

\begin{figure} [ht]
   \begin{center}
   \includegraphics[width=\linewidth]{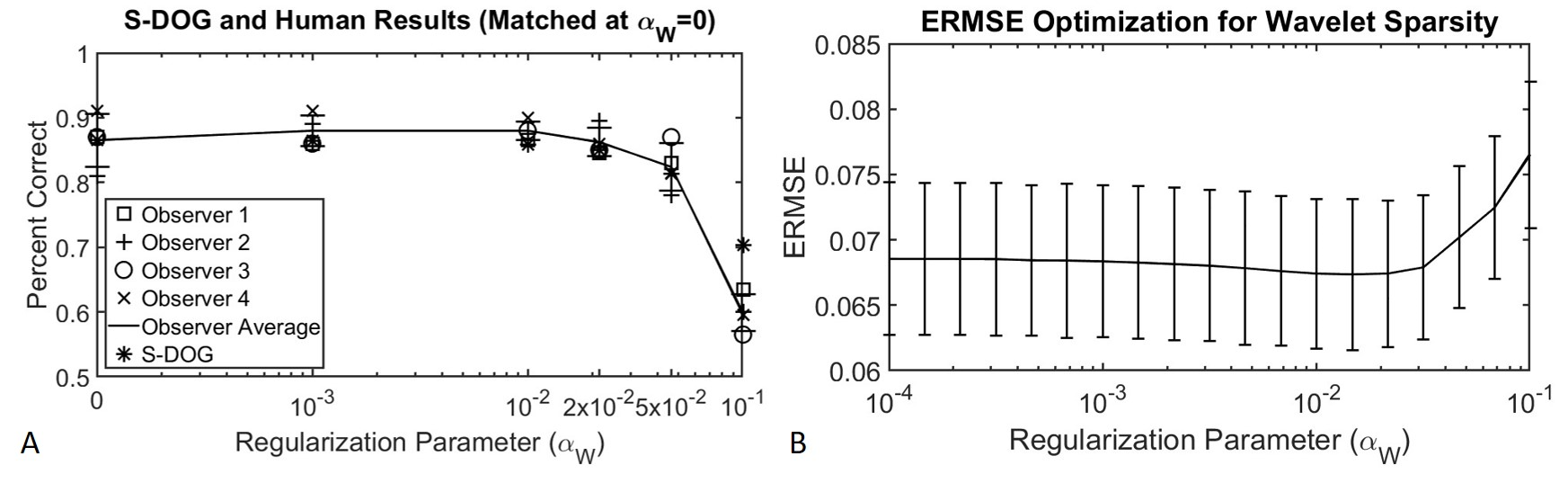}

   \end{center}
   \caption[ A) S-DOG observer with internal noise matched to average human observer for images with no wavelet sparsity.  The S-DOG observer tracked the human observer data for the other regularization parameters.  The S-DOG slightly overestimates for the largest regularization parameter.  B) Normalized ERMSE for varying wavelet sparsity.  The ERMSE metric leads to similar conclusions as the 2-AFC detection performance.] 
   { \label{fig:S-DOGfitWavelet} A) S-DOG observer with internal noise matched to average human observer for images with no wavelet sparsity.  The S-DOG observer tracked the human observer data for the other regularization parameters.  The S-DOG slightly overestimates for the largest regularization parameter. B) Normalized ERMSE for varying wavelet sparsity.  The ERMSE metric leads to similar conclusions as the 2-AFC detection performance.}
\end{figure}

\begin{figure} [H]
   \begin{center}
   \includegraphics[width=\linewidth]{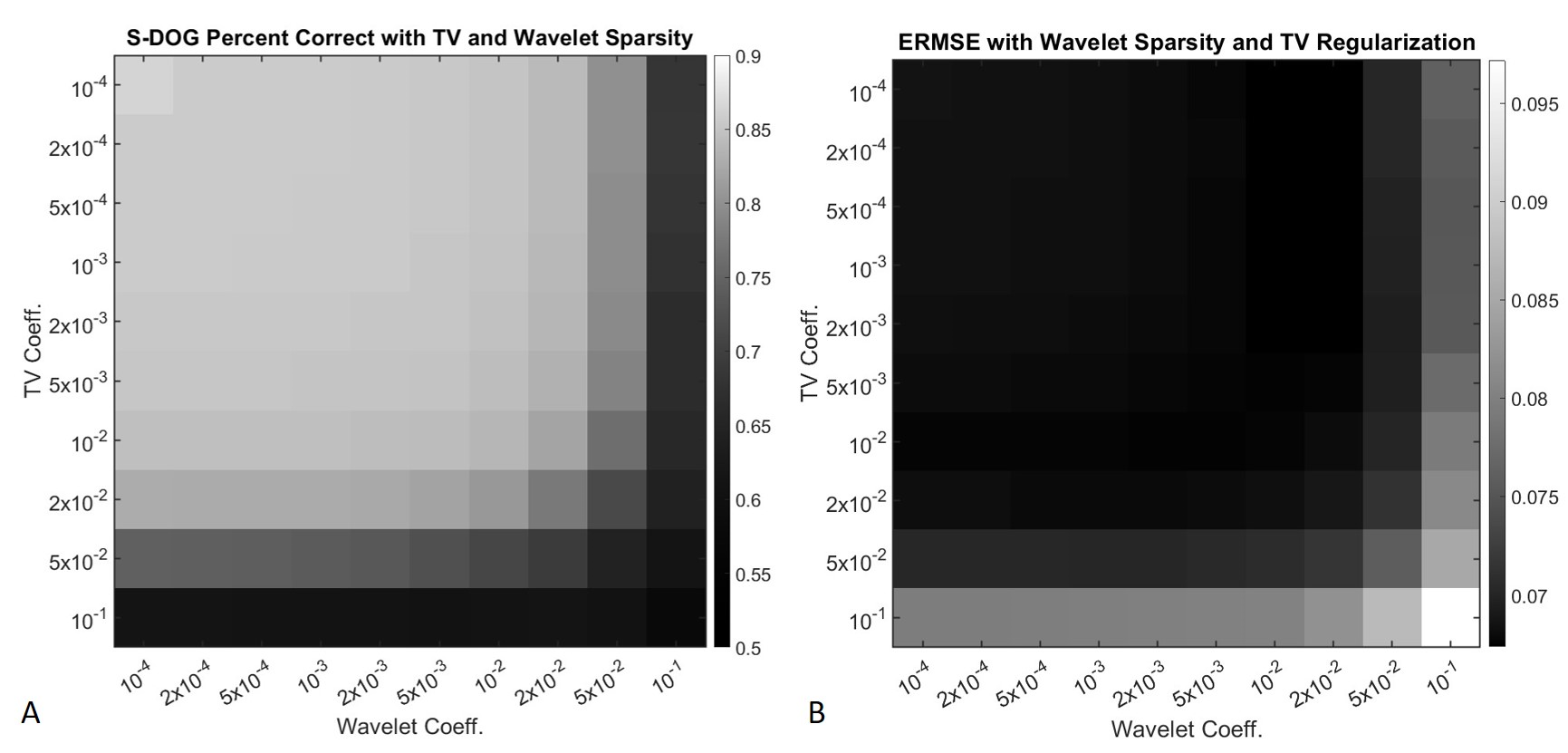}

   \end{center}
   \caption[ A) The S-DOG observer predicts that even with a combination of wavelet and TV regularization, the average human observer performance remains fairly constant for this task for a range of regularization parameters and that it degrades at high levels of regularization. B) The ERMSE behaves leads to similar conclusions as the S-DOG observer.] 
   { \label{fig:SDOGTVWfit}  A) The S-DOG observer predicts that even with a combination of wavelet and TV regularization, the average human observer performance remains fairly constant for this task for a range of regularization parameters and that it degrades at high levels of regularization. B) The ERMSE behaves leads to similar conclusions as the S-DOG observer.}
   \end{figure}

\begin{figure} [H]
   \begin{center}
   \includegraphics[width=\linewidth]{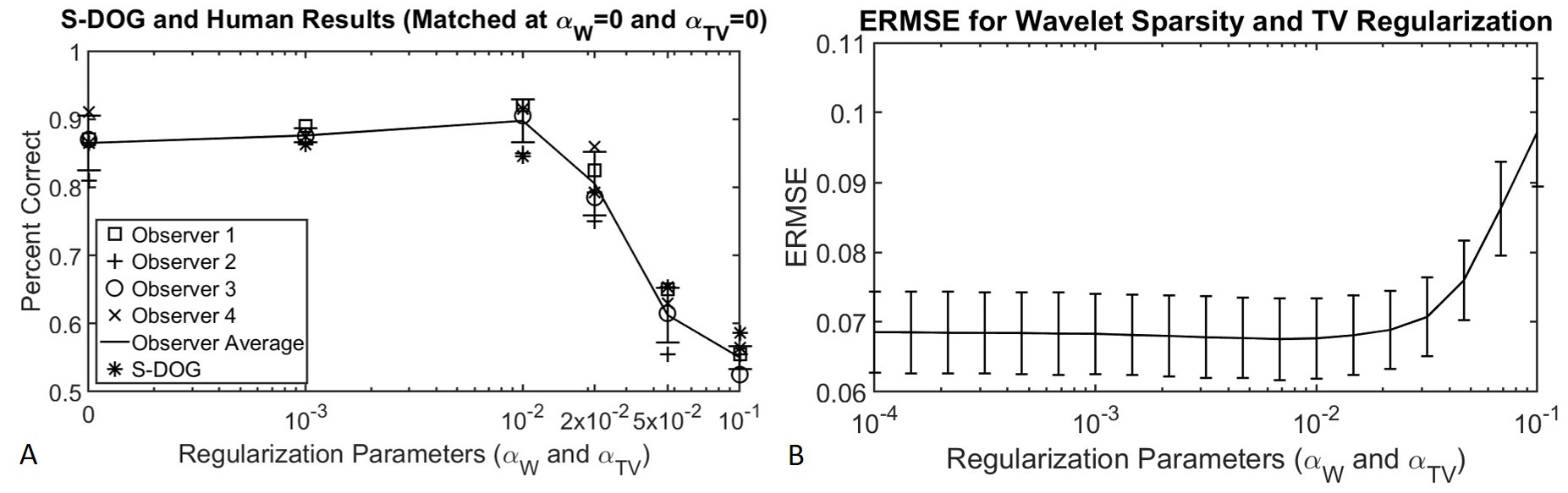}

   \end{center}
   \caption[A) The S-DOG observer predicts that even with a combination of wavelet and TV regularization, the average human observer performance remains fairly constant for this task for a range of regularization parameters and that it degrades at high levels of regularization. B) The ERMSE behaves leads to similar conclusions as the S-DOG observer.] 
   {\label{fig:SDOGTVWpred} A) The S-DOG observer predicts that even with a combination of wavelet and TV regularization, the average human observer performance remains fairly constant for this task for a range of regularization parameters and that it degrades at high levels of regularization. B) The ERMSE behaves leads to similar conclusions as the S-DOG observer.  }
   \end{figure}

Similarly, the S-DOG is also able to track human observer performance for varying wavelet regularization but again overestimates performance at the largest values of regularization as seen in Figure \ref{fig:S-DOGfitWavelet}. Wavelet regularization was not found to meaningfully improve human performance in this task. Lastly, the combination of wavelet and TV regularization does not lead to a large improvement in human performance for this detection task. Figure \ref{fig:SDOGTVWfit} shows the prediction for performance based on the S-DOG before doing the observer study where the two types of regularization were combined.    In Figure \ref{fig:SDOGTVWpred}, the predictions were validated using an observer study.  The same S-DOG model was used in this plot as in Figures \ref{fig:S-DOGfitTV}, \ref{fig:S-DOGfitWavelet}, \ref{fig:SDOGTVWfit}.  The ERMSE showed a similar behavior for all these types of regularization.  In all cases, the internal noise was chosen so that the model observer matched the average human observer performance with no regularization.  

\begin{figure} [ht]
   \begin{center}
   \begin{tabular}{c} 
   \includegraphics[width=0.9\linewidth]{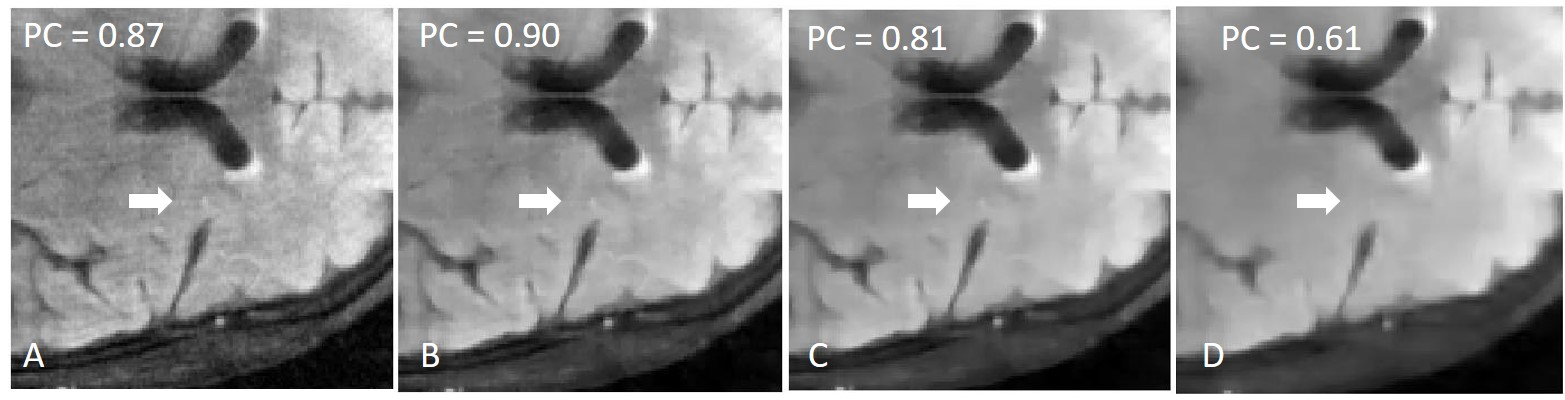}
   \end{tabular}
   \end{center}
   \caption[Sample subimages with lesions undersampled images reconstructed with TV and Wavelet regularization, A) $\alpha_{TV}=0.0,\alpha_{W}=0.0$ B) $\alpha_{TV}=0.01, \alpha_{W}=0.01$, C) $\alpha_{TV}=0.02, \alpha_{W}=0.02$, D) $\alpha_{TV}=0.05, \alpha_{W}=0.05$.  As the regularization increases, there are reduced aliasing artifacts but also reduced resolution.  Image quality and task performance (as measured by the average percent correct shown in the images) is difficult to assess from the sample images.] 
   { \label{fig:W_TVImages} Sample subimages with lesions undersampled images reconstructed with TV and Wavelet regularization, A) $\alpha_{TV}=0.0,\alpha_{W}=0.0$ B) $\alpha_{TV}=0.01, \alpha_{W}=0.01$, C) $\alpha_{TV}=0.02, \alpha_{W}=0.02$, D) $\alpha_{TV}=0.05, \alpha_{W}=0.05$.  As the regularization increases, there are reduced aliasing artifacts but also reduced resolution.  Image quality and task performance (as measured by the average percent correct shown in the images) is difficult to assess from the sample images.
}
   \end{figure}

It is difficult to quantify task performance based on subjective evaluation of image quality by looking at the images.  Figure \ref{fig:W_TVImages} show sample images with a signal for different levels of regularization.  Previous work showed that the MSE and SSIM may change in a meaningful way with regularization \cite{PinedaPMB2021} but the task performance and visual assessment remain similar.

This study considered undersampling at a high SNR.  The effect of regularization on detection performance for with varying backgrounds could be different at lower SNR regimes.  Some slight improvement was seen in the context of ramp-spectrum noise for human observers \cite{abbey2001human} and in ideal observer performance in undersampled MRI \cite{PinedaPMB2021}. Recent work studying the effect of denoising using neural networks on detection performance in SPECT showed that denoising could decrease detection performance even if it improved RSME and SSIM \cite{Yu2022}.  In another study using simulated images showed similar results \cite{Li2021}. Exploring applications where regularization helps in a detection-based perspective would be useful to better understand the regimes where metrics like ERMSE and detection performance agree and disagree.

\begin{figure} [ht]
   \begin{center}
   \begin{tabular}{c} 
   \includegraphics[width=0.9\linewidth]{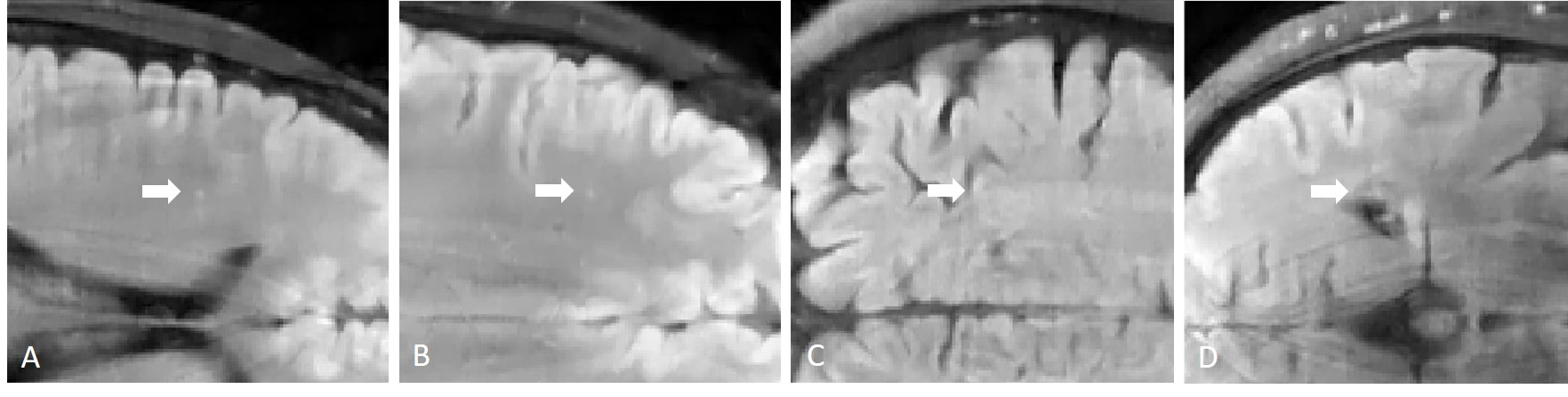}
   \end{tabular}
   \end{center}
   \caption[Sample subimages with lesions undersampled images reconstructed with TV and Wavelet regularization ($\alpha_{TV}$=0.01,$\alpha_W$=0.01).  Even though the signals are always present, subjectively we see that the signal is easier to detect in images A and B but harder to see in images C and D.  This variability in visibility due to background variability for the same reconstruction is one of the reasons why it is important to average over multiple backgrounds to estimate performance.] 
   { \label{fig:W_TVImagesOneReg} Sample subimages with lesions undersampled images reconstructed with TV and Wavelet regularization ($\alpha_{TV}$=0.01,$\alpha_W$=0.01).  Even though the signals are always present, subjectively we see that the signal is easier to detect in images A and B but harder to see in images C and D.  This variability in visibility due to background variability for the same reconstruction is one of the reasons why it is important to average over multiple backgrounds to estimate performance.
}
   \end{figure}

The visibility of the signal varies depending on the background. Figure \ref{fig:W_TVImagesOneReg} shows the variability in the sample images for a single regularization value.

A limitation of this study is the experimental design which paired the images in the 2-AFC study randomly for each of the observers.  This leads to complicated correlations within the cases and readers since readers may share the signal image in a 2-AFC trial but with different background images, for example.  To our knowledge, current approaches for unbiased estimates of variance do not include such an experimental design \cite{Gallas2009, Smith2020}.  
   
\section{Conclusion}
To our knowledge, this is the first application of a model observer for tracking human observer detection performance in undersampled MRI reconstruction.  The S-DOG model observer was able to track human performance for TV and wavelet constraints, as well as the combination of both TV and wavelet. One of the results of this study is that model observers are able to track human observer performance as the regularization changes.  We also observed a trend that commonly used regularization parameters of TV, wavelet, and combination TV and wavelet constraints led to a plateau in detection performance for low levels of regularization before causing a drop in performance for large levels of regularization.  This suggests that within that plateau, the regularization parameter could be chosen by some other criteria without affecting task performance.

\subsection*{Data, Materials and Code Availability}
The code used for placing the signals in the raw k-space data, and running the observer studies and model observers can be found at:
\noindent \url{https://github.com/MoMI-Manhattan-College/MRI-Signal-Detection}

\subsection* {Acknowledgements}
This work was supported by the National Institute of Biomedical Imaging and Bioengineering of the National Institutes of Health under award number R15-EB029172, the Manhattan College Faculty Development Grant and the Kakos Center for Scientific Computing. The authors thank Dr. Krishna S. Nayak at the University of Southern California and Dr. Craig K. Abbey at University of California, Santa Barbara for their time and guidance.   The authors would also like to thank the reviewers for their thoughtful comments which improved the manuscript and identified areas of future work.


\bibliography{SPIEJMI}   
\bibliographystyle{spiejour}   


{\bf Alexandra G. O'Neill} is a clinical research coordinator in the Division of Neuropsychiatry and Neuromodulation at Massachusetts General Hospital.  She received her BS in mathematics and psychology from Manhattan College in 2022.  Her research interests include modeling of human observer performance and statistical analysis in psychology.

{\bf Emely L. Valdez} is a mathematics teacher at DreamYard Preparatory High School. She received her BS in mathematics from Manhattan College. Her research interests include modeling of human observer performance.

{\bf Sajan Goud Lingala, PhD} is an Assistant Professor of Biomedical Engineering and Radiology at the University of Iowa. He received his Bachelors, Masters, PhD all in Biomedical Engineering respectively from the Osmania University (2002-06), Indian Institute of Technology (2006-08), University of Iowa (2008-14). Between 2014-2017, he completed a postdoctoral fellowship at the University of Southern California. His current research interests include rapid MRI sequence design, dynamic MRI, model based reconstruction using adaptive priors (eg. learning based). 

{\bf Angel R. Pineda} is a professor of mathematics at Manhattan College.  He completed his BS in chemical engineering from Lafayette College, his PhD in applied mathematics from the University of Arizona and his postdoctoral fellowship in the Radiology Department of Stanford University.  His research quantifies the information content of medical images from a task-based perspective.


\end{spacing}
\end{document}